\documentclass{appolb}
\usepackage{graphicx}

\newcommand{\tpz}{${}^{3\!}P_0$}

\begin{document}
\title{Unquenching and unitarising mesons in quark models and on the
lattice\thanks{Talk by G.~Rupp at Workshop ``Excited QCD 2016'',
Sintra, Portugal, May 7--13, 2017.}
}
\author{George Rupp\address{Centro de F\'{\i}sica e Engenharia de Materiais
Avan\c{c}ados, Instituto Superior T\'{e}cnico, Universidade de Lisboa,
P-1049-001, Portugal}
\\[5mm]
Eef van Beveren\address{Centro de F\'{\i}sica da UC, Departamento de
F\'{\i}sica, Universidade de Coimbra, P-3004-516, Portugal}
}
\maketitle

\begin{abstract}
Mesons with masses below their lowest OZI-allowed strong-decay thresholds
have very small widths. Thus, it is usually believed that they can be safely
treated as pure quark-antiquark bound states in spectroscopy models.
However, unitarised and coupled-channel models from decades ago already 
indicated that this may not be the case, owing to significant virtual
meson-loop contributions. Recent unquenched lattice calculations that include
two-meson interpolators besides the usual $q\bar{q}$ ones confirm the latter
conclusion, in particular for the enigmatic narrow 
$D_{s0}^\star(2317)$, $D_{s1}(2460)$, and $X(3872)$ states. \\
Here, we briefly review some predictions of some old and new quark models that
go beyond the static description of mesons, also in comparison with 
up-to-date lattice results.
\end{abstract}

\section{Introduction}
Knowledge of low-energy QCD is encoded in the observable
properties of hadrons, that is, mesons and baryons. Most importantly,
hadronic mass spectra should provide detailed information on the forces
that keep the (anti)quarks in such systems permanently confined,
inhibiting their observation as free particles. However, the prominent
factor that makes it very difficult to extract this information is the
lightness of current (anti)quarks as compared to the QCD scale
$\Lambda_{\mbox{\scriptsize QCD}}$. This complicates the simple model
of mesons and baryons as pure quark-antiquark ($q\bar{q}$) and three-quark
($qqq$) states, respectively, endowing them with a flurry of $q\bar{q}$
pairs being constantly created and annihilated in the strong QCD fields. \\
\indent
Now, in so-called unquenched lattice QCD, effects of $q\bar{q}$ loops are
fully taken into account, by including dynamical quark degrees of freedom
via a fermion determinant. Nevertheless, allowing for virtual $q\bar{q}$
pairs does not paint a complete picture, as the created quark and antiquark
may recombine with the  original (anti)quarks so as to form two new
colourless hadrons. Even if the mass of the initial hadron is smaller
than the sum of the new hadrons' masses, so that no real decay can take
place, the corresponding virtual processes via meson-meson or meson-baryon
loops will contribute to the total mass. This is expected to be all the more
significant according as the decay-threshold mass lies closer to the
original hadron's mass. On the other hand, if the latter mass is above
threshold, the hadron actually becomes a resonance, whose properties are
determined by $S$-matrix analyticity and unitarity. \\
\indent
In very recent years, different lattice groups have managed to extract
scattering phase shifts and resonance properties from unquenched
finite-volume simulations including meson-meson or meson-baryon
interpolating fields, besides the usual $q\bar{q}$ or $qqq$ ones,
respectively (see Ref.~\cite{ARXIV170606223} for a review hot off the
press). Some of these works on mesonic resonances \cite{PDG2016} show that
large mass shifts may result from unitarisation, even when analytically
continued to underneath the lowest strong-decay threshold. Typical examples
are the $D_{s0}^\star(2317)$ \cite{PRL111p222001} and $D_{s1}(2460)$
\cite{PRD90p034510} open-charm mesons, as well as the mysterious $X(3872)$
\cite{PRD92p034501} charmonium-like state (see Sec.~4). \\
\indent
Most quark models have been paying little or no attention to these lattice
results, despite their far-reaching implications for meson spectroscopy.
However, almost four decades ago pioneering work was already published on
coupled-channel and fully unitarised models of mesons, with predictions some
of which only now are starting to be supported in lattice QCD computations.
In the present short paper, a selection of such predictions will be reviewed,
alongside more recent model results and lattice confirmations. The
organisation is as follows: in Sec.~2 we briefly discuss the
concepts of unquenching and unitarisation in the context of quark models as
well as on the lattice. Section~3 reviews our old model predictions for the
light scalar mesons vis-\`{a}-vis very recent lattice results.
In Sec.~4 we compare our much more recent model descriptions of a few puzzling
mesonic states with the corresponding lattice calculations. Conclusions are
drawn in Sec.~5. 
\section{Unitarisation and unquenching}
\label{unquench}
The notion that the instability of most hadrons must have implications
for their spectra dates back to the late 1970s and early 1980s. The point is
that in many cases hadronic decay widths are of the same order of magnitude
as the average level splittings \cite{PDG2016}. The corresponding baryon and
meson resonances are characterised by poles in the complex-energy plane, 
whose locations are governed by $S$-matrix properties like unitarity and
analyticity. To pretend that the real parts of these pole positions should
correspond to the real energy levels of a pure confinement spectrum 
without the possibility of decay is not only naive but even a denial of
elementary scattering theory. This was realised by the 
unitarised meson models of the Cornell, Helsinki, and Nijmegen groups, with
first applications to charmonium \cite{PRD17p3090}, light pseudoscalar and
vector mesons \cite{AOP123p1,ZPC5p205}, and heavy quarkonia \cite{PRD21p772}
as well as pseudoscalar and vector mesons \cite{PRD27p1527}, respectively.
Although the quantitative
predictions for especially mass shifts turned out to be quite varied
\cite{APPS9p653}, dependent on details of wave functions as well as decay
mechanisms and included channels, these early results unmistakably showed
very significant deformations of pure confinement spectra. Even more dramatic
was the description of light scalar mesons \cite{PDG2016} as dynamical
resonances \cite{ZPC30p615} in the model of Ref.~\cite{PRD27p1527},
appearing as extra, low-lying states alongside the regular scalar mesons with
masses above 1.3 GeV. We shall come back to this work in Sec.~3. \\
\indent
Although these early meson models require the creation of $q\bar{q}$ pairs in
order to allow decay, they do not treat the quark degrees of freedom 
dynamically, using instead constituent quark masses and a phenomenological
decay mechanism, based on e.g.\ the \tpz\ model (see Ref.~\cite{PRD27p1527} and
references therein). The first prominent approaches employing light current
quarks with dynamical chiral-symmetry breaking were developed by the Orsay
\cite{PRD31p137} and Lisbon \cite{PRD42p1611} groups. In the latter work,
vacuum condensation
of \tpz\ light Dirac $q\bar{q}$ pairs due to a pure vector-like confining
potential gives rise to dynamical chiral-symmetry breaking. Through the 
subsequent yet consistent solution of a single-quark mass-gap quation, a 
$q\bar{q}$ Salpeter equation, and a meson-meson scattering equation using
the Resonating Group Method, reasonable masses and widths were obtained
for the $\rho$ and $\phi$ resonances. This was the first --- and to our
knowledge so far the only --- quark model to consider both unitarisation
and unquenching, in the sense that the quarks were treated dynamically.
Besides accounting for a massless pion in the chiral limit and a correct
low-energy $\pi\pi$ amplitude, a very important result for the present
discussion was a negative mass shift of the order of 300 MeV for the $\rho$ 
meson with respect to its confinement-only mass, owing to the coupling to the
$\pi\pi$ decay channel. This shift is
comparable to the one found before in Ref.~\cite{PRD27p1527}. However,
no further spectroscopy calculations were done in the full model of
Ref.~\cite{PRD42p1611}. On the other hand, momentum-space versions of the
Nijmegen model \cite{PRD21p772,PRD27p1527,ZPC30p615}, originally formulated
in  coordinate space, have been applied to a variety of light, heavy-light,
and heavy mesons in more recent years. A few typical examples will be reviewed
in Sec.~4.
\\ \indent
The term ``unquenched'' first appeared in the title of a lattice paper due to
I.~Montvay \cite{PLB139p70} back in 1984, describing a Monte Carlo calculation
with virtual quark loops through a quark determinant for Wilson fermions.
Progressively, many other lattice groups started to carry out QCD simulations
with dynamical quarks. However, only much later hadrons were allowed to decay
on the lattice, by employing L\"{u}scher's finite-volume method
\cite{NPB354p531} and generalisations thereof \cite{ARXIV170606223} to extract
scattering phase shifts from discrete energy levels for different lattice
sizes. Also in quark models the term ``unquenching'' is being used more and
more, although it usually is a sloppy way of referring to some unitarisation
or coupled-channel approach. Two examples are a meson toy model 
\cite{JPG31p845} and the baryon model of Ref.~\cite{PRC80p065210}.
\section{Light scalar mesons}
\begin{figure}[!h]
\begin{center}
\includegraphics[trim = 40mm 175mm 50mm 45mm,clip,width=10cm,angle=0]
{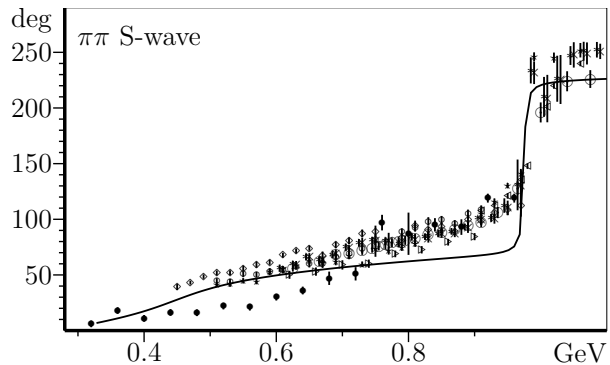}
\end{center}
\caption{$S$-wave $\pi\pi$ phases predicted in Ref.~\cite{ZPC30p615}
(see text and reference for data).}
\label{pipi}
\end{figure}

In Ref.~\cite{ZPC30p615} a complete light scalar-meson nonet emerged, as
additional and dynamical resonances, by
employing the unitarised multichannel model of Ref.~\cite{PRD27p1527} with
unaltered parameters. Now, more than 30 years later, those predictions for
masses, widths, and pole positions are still fully compatible with
experiment \cite{PDG2016}. In Fig.~\ref{pipi} we show the computed
$S$-wave $\pi\pi$ phase shift, obtained without any fit, together with the
then available data. \\ \indent
Very recent lattice calculations
\cite{PRL118p022002,PRD91p054008,PRD93p094506} confirm our original
\cite{ZPC30p615} interpretation of the light scalars as $P$-wave $q\bar{q}$
states with large meson-meson components. In particular,
Ref.~\cite{PRL118p022002} described the lightest isoscalar scalar meson
in a lattice calculation with $q\bar{q}$ and $\pi\pi$ interpolating fields,
though with still too large pion masses, viz.\ of 236 and 391~MeV.
Nevertheless, in the former case a broad $\sigma$-like (cf.\ $f_0(500)$
\cite{PDG2016}) resonance shows up, which becomes a bound state for the
larger $\pi$ mass. In the isodoublet case, the same lattice group studied
\cite{PRD91p054008} --- among others --- a light scalar state, using
$q_1\bar{q}_2$ (light-strange combination), $\pi K$, and $\eta K$
interpolators, and found a virtual bound state for a pion mass of 391~MeV.
This is expected \cite{PRD91p054008} to evolve into a broad resonance for
a more physical pion mass, thus being a strong candidate for $K_0^\star(800)$
\cite{PDG2016} (``$\kappa$''). Finally, again the same collaboration found
\cite{PRD93p094506} a scalar isovector resonance in a calculation with
$q\bar{q}$, $\pi\eta$, and $K\bar{K}$ interpolating fields, representing
most likely $a_0(980)$ \cite{PDG2016}.
\section{Very narrow $D_{s0}^\star(2317)$, $D_{s1}(2460)$, and $X(3872)$ mesons}
In Ref.~\cite{PRL91p012003} we applied a very simple momentum-space version
of the unitarised model employed in Refs.~\cite{PRD21p772,PRD27p1527,ZPC30p615}
to a scalar $c\bar{s}$ state  strongly coupled to the $S$-wave $DK$ 
channel. This allowed to explain the just discovered puzzling and very narrow
$D_{s0}^\star(2317)$ \cite{PDG2016}. In Ref.~\cite{PRL111p222001} a lattice
calculation with $c\bar{s}$ and $DK$ interpolating fields confirmed our
interpretation. The same lattice group also studied \cite{PRD90p034510} the
narrow axial-vector charm-strange meson $D_{s1}(2460)$ \cite{PDG2016} besides
other $D_s$ states, finding a bound state below the $D^\star K$ threshold, thus
confirming our findings \cite{PRD84p094020} in a multichannel generalisation of
the momentum-space model used in Ref.~\cite{PRL91p012003}. Last but not least,
again the same lattice collaboration investigated the controversial
$J^{PC}=1^{++}$ charmonium-like meson $X(3872)$, in various simulations with
$c\bar{c}$, tetraquark, and $D^\star D$ interpolating fields. This study was
particularly interesting, because it also searched for signals of tetraquark
charmonium-like states up to 4.2~GeV, not finding any. As for $X(3872)$, this
state only survived with both $c\bar{c}$ and $D^\star D$ interpolators
included, the tetraquark ones being largely irrelevant. This lends support to
our unitarised description \cite{EPJC71p1762} of $X(3872)$ with a $c\bar{c}$
component coupled to several open-charm meson-meson channels.
\section{Conclusions}
In the present short review we have tried to convey the message that modern
meson spectroscopy must take both real and vitual strong decay into account
in order to arrive at minimally reliable predictions. In particular, when
a meson is narrow, one cannot automatically conclude that coupled-channel
effects will also be small, since a (quasi-)bound state may result from a
negative mass shift due to meson-meson loops, driving a bare $q\bar{q}$
state  to below its lowest threshold. Also in such cases $S$-matrix unitarity
and analyticity should serve as a guidance. Several recent lattice
calculations leave no doubt that the unitarisation issue should finally be
taken seriously.

\end{document}